\documentclass[twocolumn,aps,prl,amsmath,showpacs]{revtex4-1}
\pdfoutput=1
\usepackage{graphicx,dcolumn,bm,amssymb,amsmath}
\usepackage{color}

\usepackage{float}

\usepackage{hyperref}
\def\be{\begin{equation}}
\def\ee{\end{equation}}

\begin{document}

\preprint{}[IFT-UAM/CSIC-19-60]

\title{\Large Hydrodynamics of disordered marginally-stable matter}

\author{Matteo Baggioli}%
\email{matteo.baggioli@uam.es}
 
\affiliation{Instituto de Fisica Teorica UAM/CSIC, c/Nicolas Cabrera 13-15,
Universidad Autonoma de Madrid, Cantoblanco, 28049 Madrid, Spain.}
\author{Alessio Zaccone}
\email{alessio.zaccone@unimi.it}
\affiliation{Department of Physics ''A. Pontremoli", University of Milan, via Celoria 16, 20133 Milan, Italy.}

\affiliation{Department of Chemical Engineering and Biotechnology,
University of Cambridge, Philippa Fawcett Drive, CB30AS Cambridge, U.K.}
\affiliation{Cavendish Laboratory, University of Cambridge, JJ Thomson
Avenue, CB30HE Cambridge, U.K.}
\begin{abstract}
\noindent 
We study the vibrational spectra and the specific heat of disordered systems using an effective hydrodynamic framework. We consider the contribution of diffusive modes, \textit{i.e.} the 'diffusons', to the density of states and the specific heat. We prove analytically that these new modes provide a constant term to the vibrational density of states $g(\omega)$. This contribution is dominant at low frequencies, with respect to the Debye propagating modes. We compare our results with numerical simulations data and random matrix theory. Finally, we compute the specific heat and we show the existence of a linear in $T$ scaling $C(T) \sim  c\,T$ at low temperatures due to the diffusive modes. We analytically derive the coefficient $c$ in terms of the diffusion constant $D$ of the quasi-localized modes and we obtain perfect agreement with numerical data. The linear in $T$ behavior in the specific heat is stronger the more localized the modes, and crosses over to a $T^{3}$ (Debye) regime at a temperature $T^{*}\sim \sqrt{v^{3}/D}$, where $v$ is the speed of sound. Our results suggest that the anomalous properties of glasses and disordered systems can be understood effectively within a hydrodynamic approach which accounts for diffusive quasi-localized modes generated via disorder-induced scattering. 
\end{abstract}


\maketitle

Glasses and disordered systems display interesting anomalous properties which remain unexplained to date. The vibrational modes, the thermodynamic quantities and the thermal transport coefficients strongly differ from the long known results for ordered crystals and an underlying robust theoretical picture is still absent~\cite{Kob}.

A widely accepted view is that the vibrational degrees of freedom in glasses are not simply and only the standard Debye propagating phonons which obey, at low energy, the known dispersion relation:
\begin{equation}
\omega_{T,L}\,=\,v_{L,T}\,k\,+\,\dots
\end{equation}
where $v_{T,L}$ is the speed of propagation of the transverse (T) and longitudinal (L) modes.
\begin{figure}[hbt]
\centering
\includegraphics[width=0.7\linewidth]{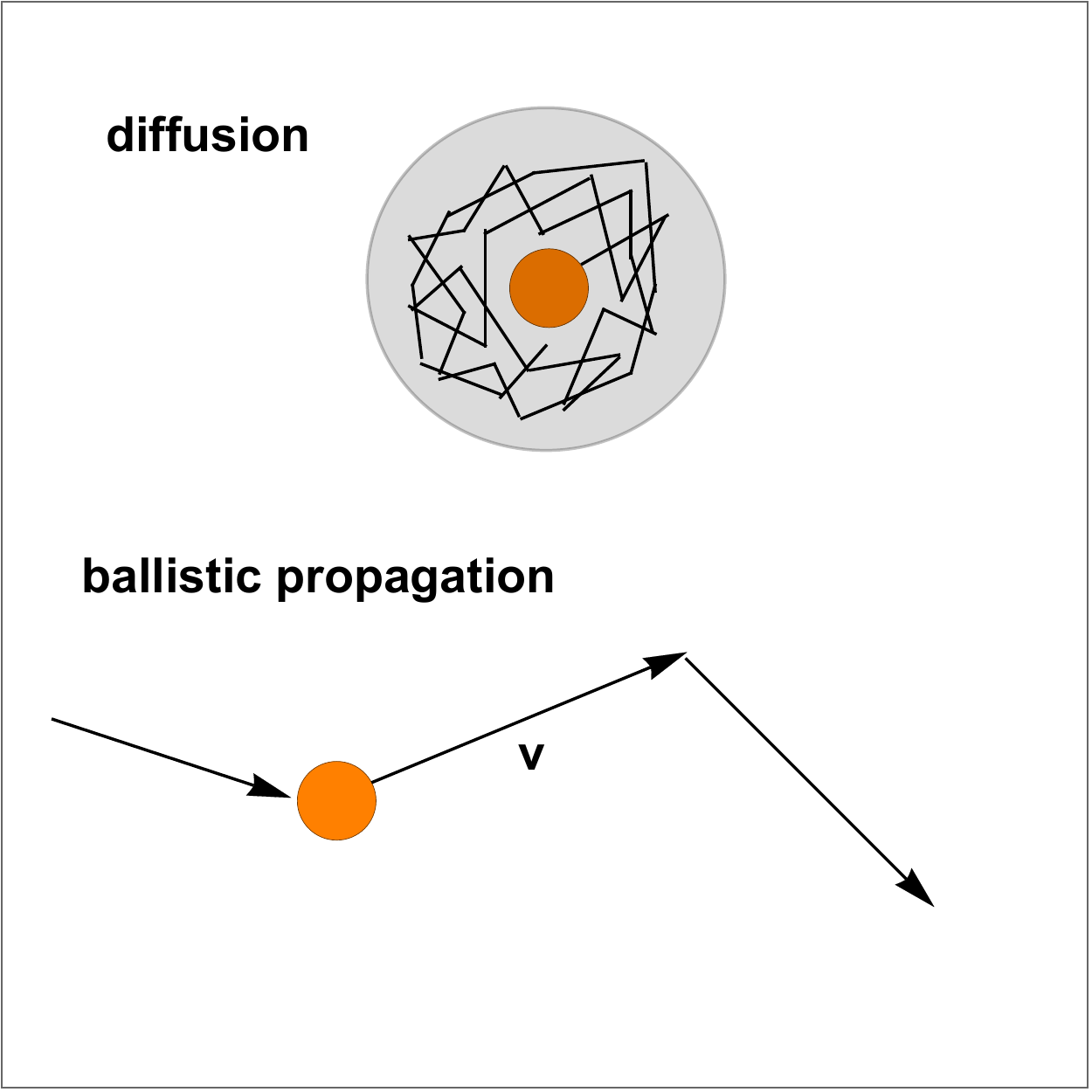}

\vspace{0.5cm}

\includegraphics[width=0.7\linewidth]{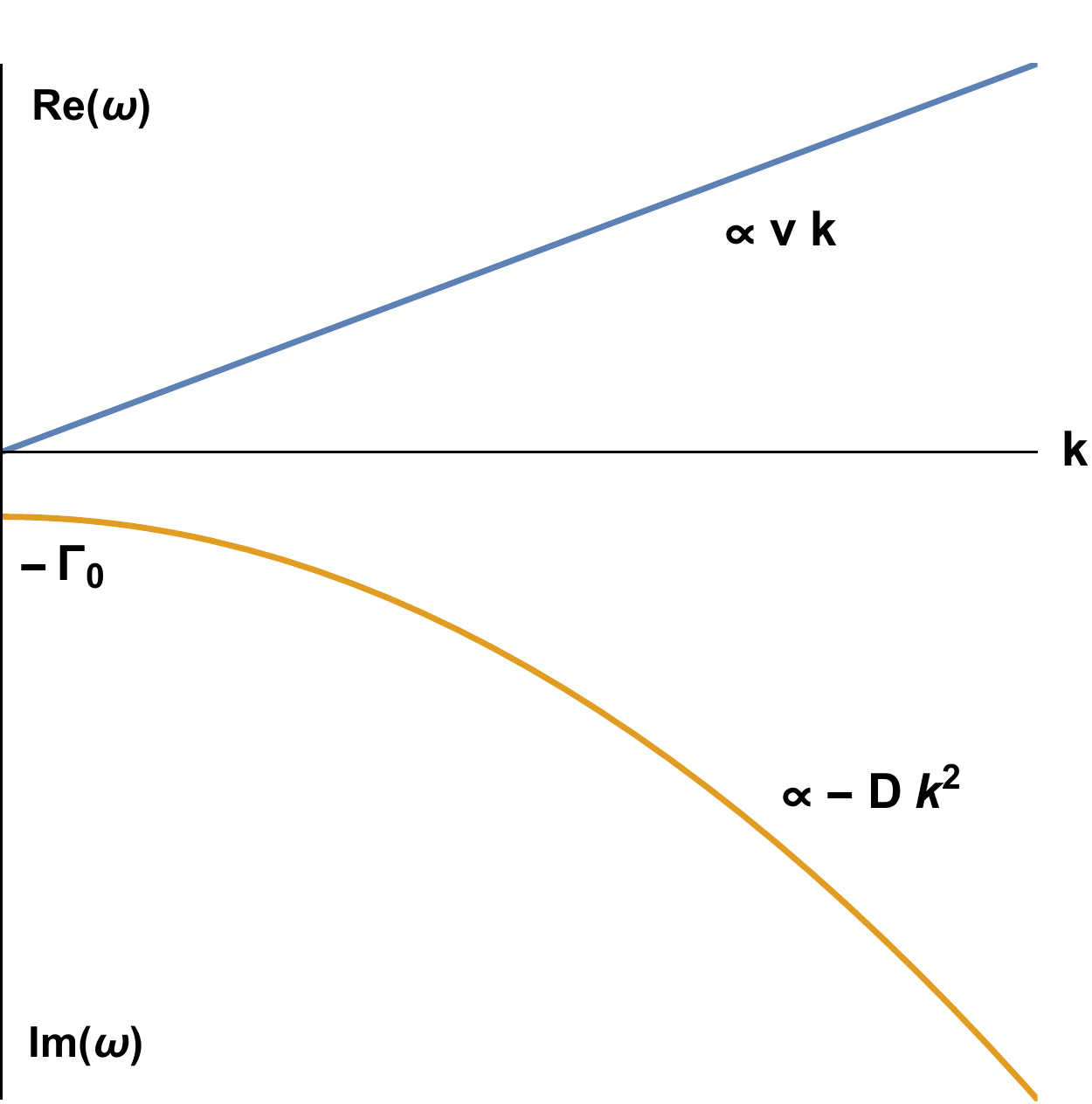}

\caption{A visual representation of the difference between diffusive dynamics and ballistic propagation. \textbf{Top: }A cartoon of the diffusive and ballistic mechanisms. \textbf{Bottom: }The dispersion relations for a propagating mode $\omega=v \,k$ and a damped diffusive mode $\omega=-i \,\Gamma_0 -i \,D \,k^2$.}
\label{fig0}
\end{figure}\\
One possible extension of Debye's theory involves the introduction of low-energy diffusive excitations which appear due to the strong disorder typical of glassy systems~\cite{allen1999diffusons}. The presence of such modes can be understood, from a physical point of view, as a consequence of frequent scattering events that are caused by the absence of long-range order~\cite{allen1999diffusons,Schirmacher} (see Fig.\ref{fig0} for a visual representation), and possibly promoted at the microscopic level by the lack of inversion symmetry~\cite{Milkus} and, at higher temperatures, by anharmonicity~\cite{Akkermans}. More specifically, the diffusons cannot propagate over long distance as they have a wavelength comparable with their mean-free path (Ioffe-Regel), and they appear as quasilocalized excitations.

The diffusons follow a Brownian motion dynamics which is expressed via the standard relation for the mean displacement squared:
\begin{equation}
\langle r^2 \rangle \,=\,D\,t.
\end{equation}
Here $D$ is the macroscopic diffusion constant which arises from the collective behaviour of such quasilocalized modes. The previous expression has to be contrasted with the ballistic dynamics
\begin{equation}
\langle r^2 \rangle \,=\,v^2\,t^2
\end{equation}
which is typical of the propagating phonon modes in ordered crystals. In other words, the propagating sound modes are now coexisting with additional diffusive hydrodynamic degrees of freedom.  When the diffusion constant of the diffusons becomes zero, as it happens at the mobility edge, they become completely localized and they are sometimes denominated \textit{locons}. In disordered solids this type of Anderson-localization is typically observed near the upper limit of the vibrational density of states (VDOS)~\cite{Vitelli}.

The idea that diffusons are present and extremely relevant in glasses has been introduced in \cite{allen1999diffusons} and hinted in several theoretical and experimental works in the past \cite{PhysRevLett.117.045503,Beltukov2011,PhysRevB.46.6131,PhysRevE.79.021308,PhysRevE.81.021301,Vitelli,grigera2002vibrational}.

Moreover, the onset of vibrational diffusion relates with two important topics in the physics of disordered systems: the description of disordered solids in terms of random matrix theory \cite{ParisiVDOS,Manning1,Manning2,Franz14539,parisi2014soft,benetti2018mean} and the excess of vibrational degrees of freedom at the Ioffe-Regel crossover \cite{PhysRevB.61.12031}.

In this letter we build a hydrodynamic model for the diffusons, described as collective excitations with a diffusive dispersion relation, and study their contribution to the VDOS and the specific heat of the system using an effective field theory (EFT) approach. The diffusons are shown to produce a constant-in-frequency contribution to the VDOS which dominates at low frequencies over the standard Debye term $g_{\text{Debye}}(\omega)\sim \omega^2$. Moreover, such contribution produces a linear in $T$ term in the low temperature specific heat of the system.

Importantly, we will draw interesting connections with the results recently obtained via simulations and random matrix theory in \cite{baggioli2019random}, where similar effects have been observed. These connections hint at a deeper and more fundamental link between random matrix statistics of eigenvalues and eigenmode quasilocalization and diffusion.\\
To summarize, our results suggest that beyond the two-level states (TLS) theory \cite{Phillips1,Phillips2,anderson1972anomalous} there are other possible theoretical explanations for the linear in $T$ specific heat in glasses and amorphous materials which do not rely on any  quantum phenomenon. With the present letter we provide an effective alternative explanation for such observation, which relies simply on hydrodynamic arguments and the scattering-induced diffusive nature of the vibrational degrees of freedom at low frequencies, \textit{e.g.} the diffusons.
 
%
We consider collective vibrational modes whose dispersion relation has an imaginary part. In particular we discuss collective excitations defined by the following dispersion relation:
 \begin{equation}
 \omega\,=v\,k\,-\,i\,\Gamma(k)\,=vk\,-\,i\,\Gamma_0\,-\,i\,D\,k^2\,+\,\mathcal{O}(k^4)\label{locdisp}
 \end{equation}
 in terms of a generic damping coefficient $\Gamma$ which depends on the momentum $k$ of the mode. 
 This relation can be derived from the equation of elastodynamics of a solid supplemented with a viscous contribution~\cite{Chaikin}, or from macrosopic balance equations that allow for energy dissipation~\cite{Parodi}. 
 
In this work, we consider the limit where the imaginary (diffusive, dissipative) part of the above dispersion relation effectively dominates over the real part (representing the acoustic, elastic, propagative component). Effectively, this reduces to set the ballistic propagation speed to zero, $v=0$, at least for the transverse phonons. We will show that by taking this limit, the characteristic properties of marginally stable (jammed) solids~\cite{OHern} can be reproduced.
 
We use a hydrodynamic approach to express the low-momentum limit of the damping coefficient in terms of a momentum-independent term $\Gamma_0$ and a diffusion constant $D$. Higher order terms can be ignored since they will not contribute to the low energy properties of the system. In the limit $\Gamma_0 \rightarrow 0$ these excitations are purely diffusive and coincide with those labeled in the previous literature as \textit{diffusons} \cite{allen1999diffusons}.

Hence, in the limit where diffusion dominates, the above dispersion relation becomes purely imaginary
\begin{equation}
 \omega_{\text{diff}}\,=\,-\,i\,\Gamma(k)\,=\,-\,i\,\Gamma_0\,-\,i\,D\,k^2\,+\,\mathcal{O}(k^4).\label{locdisp}
 \end{equation}

 \begin{figure}[hbtp]
\centering
\includegraphics[width=0.95\linewidth]{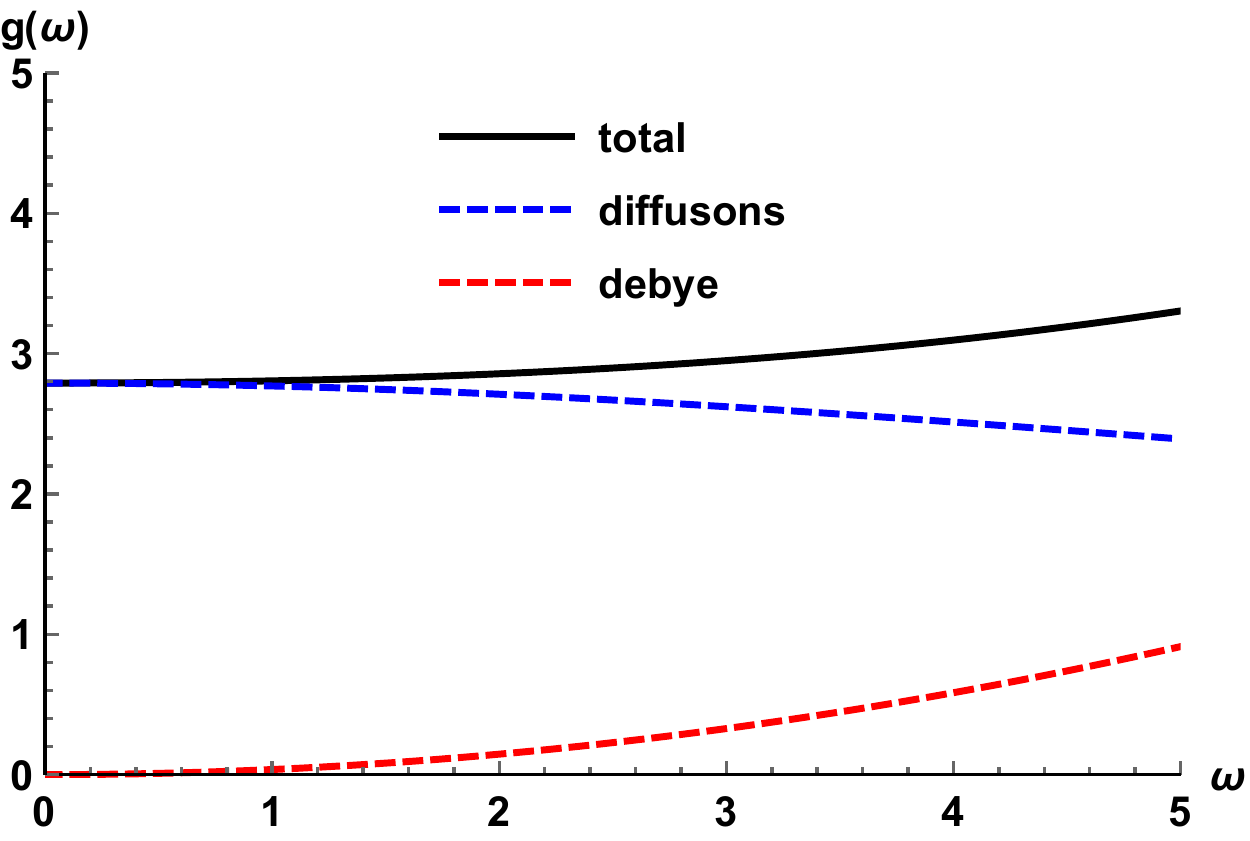}
\caption{Contributions of the diffusons and the Debye modes to the VDOS. The parameters are fixed to $V=1$,  $\Gamma_0=5$, $D=0.8$, $v_T=0.7$, $v_L=0.9$ and $k_D=5$. The qualitative results are independent of the choice of those numerical values.}
\label{fig:uno}
\end{figure}
In the opposite limit, where the imaginary part is zero, the VDOS of the Debye modes can be obtained using standard methods as
\begin{equation}
g_{\text{Debye}}(\omega)\,=\,\frac{\omega^2\,V}{2\,\pi}\,\left(\frac{2}{v_T^3}\,+\,\frac{1}{v_L^3}\right)\label{Debyedens}
\end{equation}
where $v_{L}$ and $v_{T}$ are the longitudinal and the transverse speed of sound, respectively; and $V$ is simply the volume of the system. This equation constitutes the basis of Debye theory for ordered crystals which is strongly violated in glasses and amorphous systems. In this work we will not discuss the anomaly known as boson peak (BP) which appears in glasses and amorphous systems (and also in some ordered crystals \cite{Moratalla}). The boson peak can be retrieved, within the current approach, from the competition between the real and imaginary parts in Eq. (4) as shown in \cite{baggioli2018universal,baggioli2018soft}.
Here we are interested in the dynamics of the system at lower frequencies, below the typical BP location, $\sim$ THz.

Going back to the newly introduced diffusive modes Eq.\eqref{locdisp}, we can write down their contribution to the VDOS using the formula
\begin{equation}
g_{\text{diff}}(\omega)\,=\,\frac{2\,\omega}{\,\pi\,k_{D}^{3}}\, \text{Im} \int_0^{k_D}\frac{k^2}{\omega^2\,-\,i\,\omega\,\Gamma_{0}\,-\,i\,\omega\,D\,k^2}\,\,dk
\end{equation}
where $k_D$ is the Debye momentum. The above formula can be derived by defining the VDOS as a sum of Dirac deltas centered on the eigenfrequencies and then using the Plemelj identity~\cite{Boettger}.
The integral can be performed analytically and it gives the following expression (where we omit constant prefactors):
\begin{equation}
g_{\text{diff}}(\omega)\,=\,\ \text{Re} \left[\frac{\sqrt{D} \,k_D-\sqrt{\Gamma_0 +i \,\omega} \,\tan ^{-1}\left(\frac{\sqrt{D} \,k_D}{\sqrt{\Gamma_{0} +i \,\omega}}\right)}{D^{3/2}}\right].
\end{equation}

Importantly, the contribution of the diffusons to the VDOS displays the low frequency expansion
\begin{equation}
g_{\text{diff}}(\omega)\,=\, a+ b \,\omega^2\,+\,\mathcal{O}(\omega^4)\label{lowexp}
\end{equation}
with:
\begin{equation}
a\,=\,\frac{k_D}{D}-\frac{\sqrt{\Gamma_{0} } \tan ^{-1}\left(\frac{\sqrt{D}\, k_D}{\sqrt{\Gamma_0 }}\right)}{D^{3/2}}.\label{ff}
\end{equation}
The most important observation here is that the diffusons provide a constant contribution to the VDOS, which is finite even at zero frequency and can dominate the low energy thermodynamic properties of the system. Let us notice that the dispersion relation Eq.\eqref{locdisp}, and consequently the result Eq.\eqref{ff}, are not reliable in the limit of very large diffusion constant $D \rightarrow \infty$ where higher order terms in Eq.\eqref{locdisp} have to be considered.

Finally, from the VDOS, we can obtain the specific heat of the system and in particular we can separate the contribution of the diffusons from the contribution of the propagating Debye modes using the standard formula~\cite{Khomskii}
\begin{equation}
C(T)\,=\,k_B\,\int_0^\infty \,\left(\frac{\hbar\,\omega}{2\,k_B\,T}\right)^2\,\sinh \left(\frac{\hbar\,\omega}{2\,k_B\,T}\right)^{-2}\,g(\omega)\,d\omega.
\end{equation}
\begin{figure}
\centering
\includegraphics[width=0.95\linewidth]{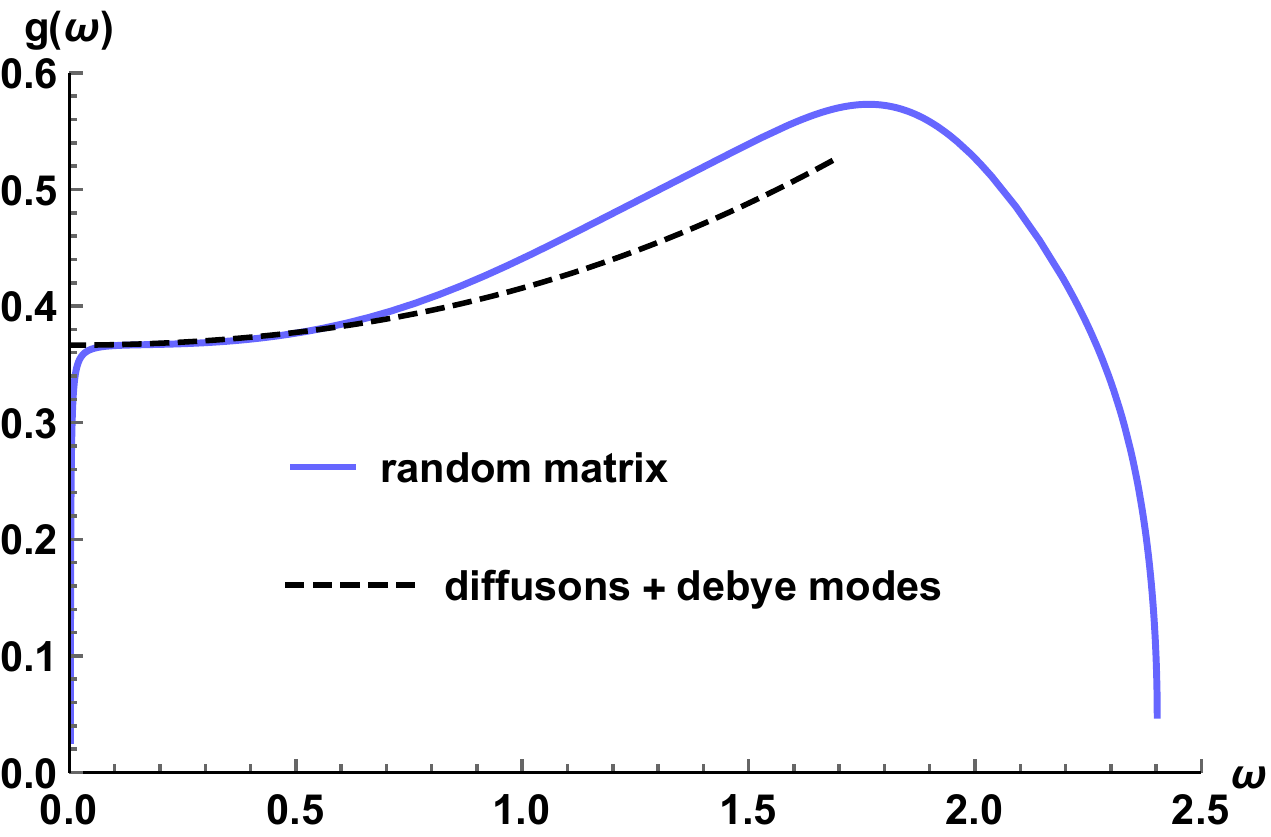}
\caption{The comparison between the VDOS for $Z=Z_c=6$ obtained via simulations of a 3D harmonic random spring network~\cite{Milkus} and best fitted by random matrix theory in \cite{baggioli2019random} (solid line) and the VDOS of the hydrodynamic model of this paper (dashed line). The simple model can reproduce the behaviour at small frequency just in terms of diffusive and (longitudinal) Debye modes. Note some discrepancy at $\omega<0.1$ due to the fact that the numerical network has $Z$ not exactly $6$ and slightly larger, whereas the theory assumes a fully marginally-stable system for which there is no going down to zero, in agreement with previous findings ~\cite{OHern}.}
\label{fig:due}
\end{figure}\\
It turns out that the diffusive modes dominate the specific heat at low temperature producing a very clear linear in $T$ contribution. In particular we have
\begin{equation}
C(T)\,=\underbrace{c\,T}_{diffusons}+\underbrace{d\,T^3}_{Debye}\,+\,\dots
\end{equation}
where the second term is the known Debye $\sim T^3$ contribution which comes from the Debye VDOS Eq.\eqref{Debyedens}, while the first term comes directly from the diffusons. Using the low frequency expansion of the density of states given in Eq.\eqref{lowexp} we can analytically derive that:
\begin{equation}
c\,=\,\frac{\pi^2}{3}\,a \label{analT}
\end{equation}
The final result gives the same linear in $T$ contribution as given by the tunnelling two-level state (TLS) theory \cite{Phillips1,Phillips2,anderson1972anomalous}, without having to resort to any quantum mechanism. 
Previous work established a relation between the anomalous thermal conductivity of glasses and the diffusons \cite{allen1999diffusons}, hence it would be appealing to provide a unifying mechanism to explain all thermal anomalies of glasses based on diffusons.
The root cause of the diffusive excitations is to be identified in many microscopic scattering processes which are taken into account here at the level of effective field theory (EFT).
As shown below, this EFT description is able to reproduce the universal linear scaling of the specific heat observed in glasses and amorphous materials.

We are now in the position of discussing the various implications of the simple hydrodynamic model introduced in the previous section. The main question is how the diffusons might modify the physical properties of the system.
\begin{figure}[hbtp]
\centering
\includegraphics[width=0.95\linewidth]{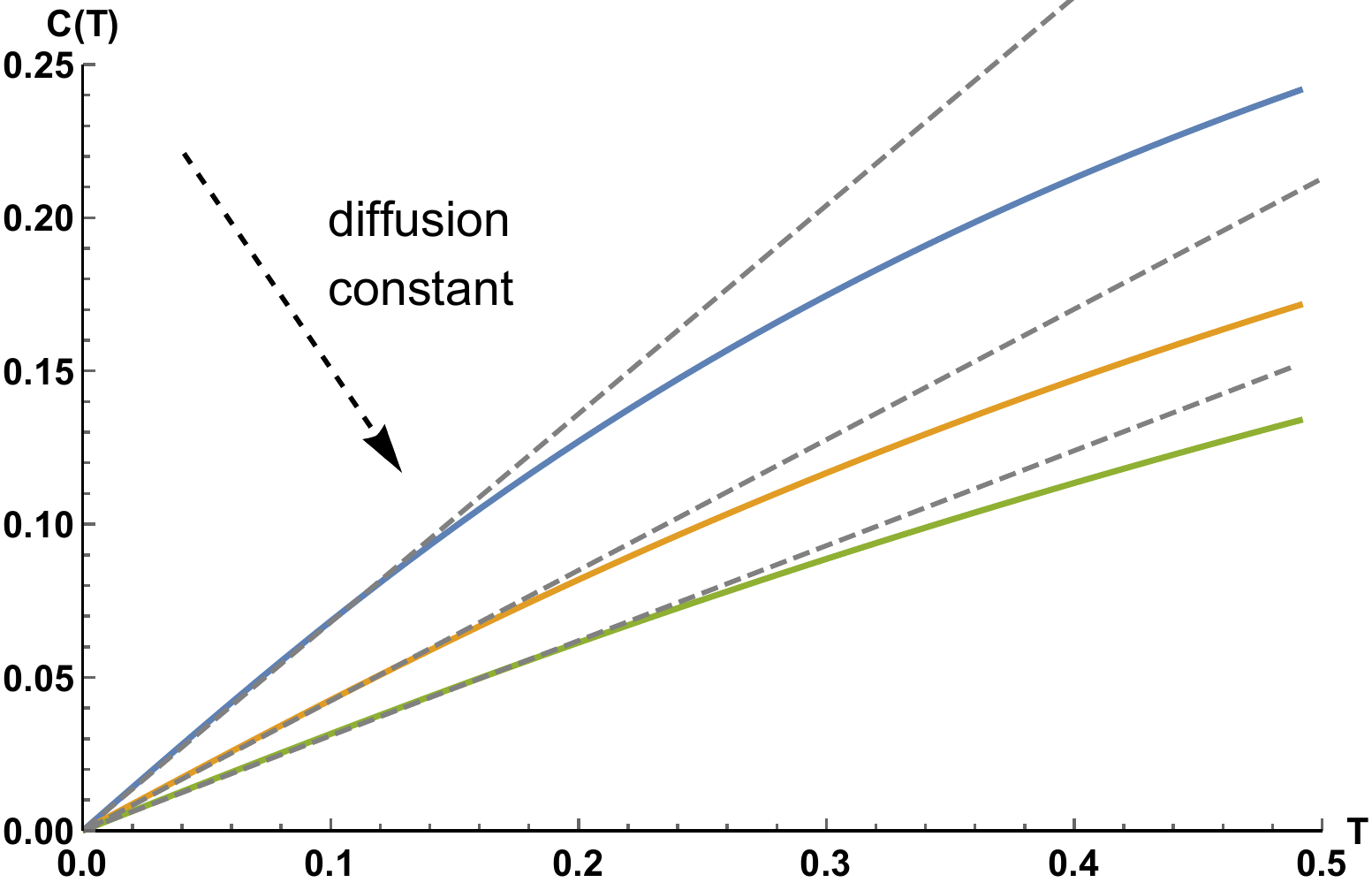}
\caption{The contribution of the diffusons Eq.\eqref{locdisp} to the specific heat for different values of the diffusion constant $D$ (increasing from blue to green). The dashed lines guide the eyes towards the linear in $T$ scaling at low temperature. The $T^3$ Debye contribution is omitted.}
\label{fig:tre}
\end{figure}\\
The first important consequence of the existence of diffusive modes of the type Eq.\eqref{locdisp} is the appearance of a constant low-frequency plateau in the total VDOS which is shown in in Fig.\ref{fig:uno}.
We have ascertained that this behaviour is qualitatively independent from the values of the parameters and it agrees with the analytical result Eq.\eqref{ff}. It is straightforward to notice that this new contribution can be dominant at low frequency with respect to the Debye term $\sim \omega^2$ (see Fig.\ref{fig:uno}), as is indeed seen in random jammed packings near the marginal stability threshold~\cite{OHern}.
\begin{figure}[hbtp]
\centering
\includegraphics[width=0.95\linewidth]{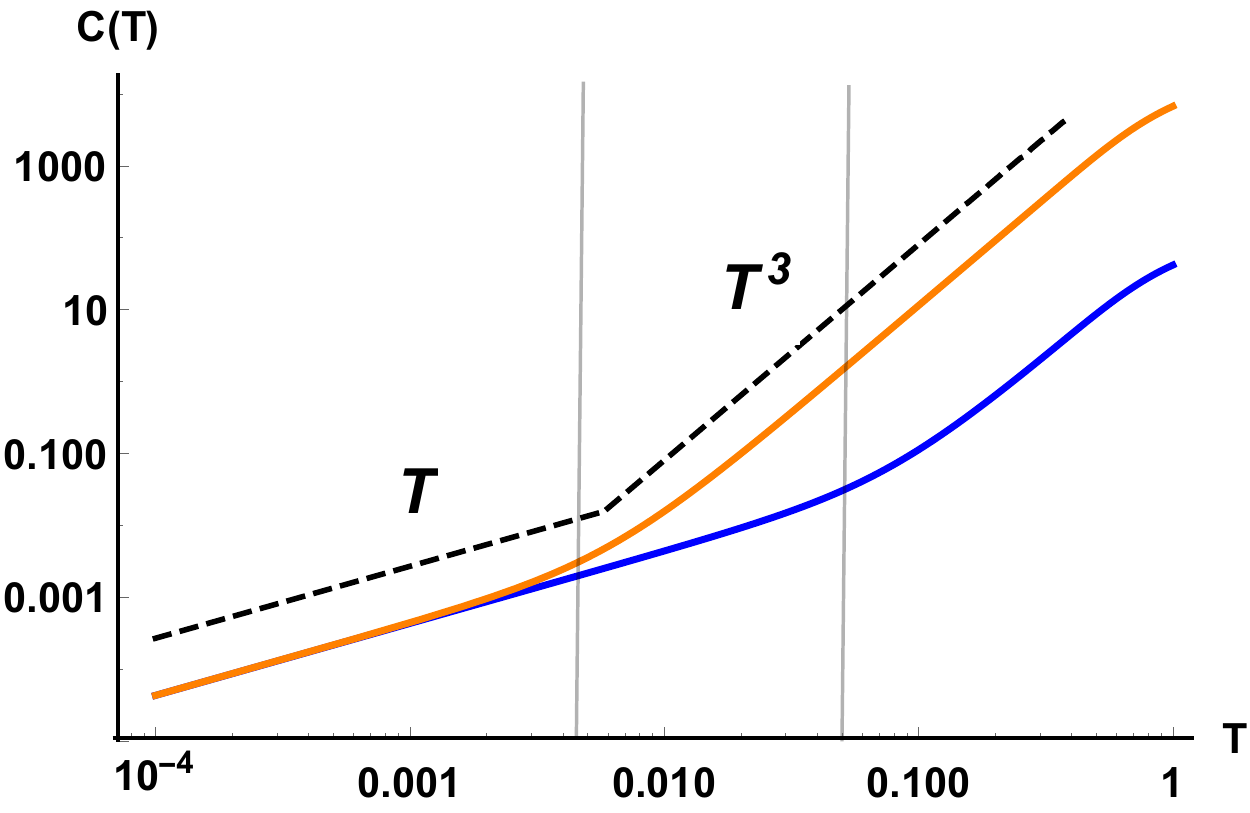}
\caption{The competition between the linear in $T$ contribution coming from the diffusons an the $T^3$ Debye term in the specific heat. The two curves have the same diffusion constant $D$ but different sound speeds for the propagating phonons. The orange curve has lower values and therefore bigger Debye contribution. The gray lines guide the eye towards the position of the crossover.}
\label{fig:new}
\end{figure}\\
Moreover, it is important to compare the results from this simple hydrodynamic model with the predictions of the simulations of random lattices and random matrix theory.  We consider the example of a random network of harmonic springs, studied in previous work~\cite{Milkus}, which features exactly the same behaviour of jammed frictionless packings at the marginally-stable limit~\cite{OHern}, both for the elastic constants and for the VDOS. The dashed line shown in Fig. \ref{fig:due} is calculated by using $g(\omega)=g_{Debye}(\omega)+g_{\text{diff}}(\omega)$, or equivalently using Eq. (7) with a Green's function in the integral which has acoustic poles coexisting with the diffusive poles.
Importantly, for the marginally-stable jammed state with $Z=6$, the transverse acoustic part is identically zero, and $v_{T}=0$, because the shear modulus vanishes at the jamming point $Z=2\,d$ with the law $G \propto (Z-2\,d)$, due to nonaffine relaxations~\cite{Zaccone2011}. 

We can observe that, at low frequencies, the behaviour is totally dominated by the diffusons contribution. In Fig. \ref{fig:due} we compare the hydrodynamic model with an analytical best fit, based on random matrix theory, of simulations data of a random network of harmonic springs at the marginal stability point $Z=Z_c =6$ \cite{baggioli2019random}. The qualitative agreement at low frequency is good and suggests the existence of a deep connection between diffusons, random matrix theory and disordered marginally-stable states. In particular, this comparison demonstrates that, at the jamming point of marginally-stable disordered solids, where the shear modulus vanishes, and only longitudinal modes propagate, the frequency spectrum is dominated by the diffusive modes (which are continuously scattered by disorder).

We then compute the contribution of the diffusive modes Eq.\eqref{locdisp} to the specific heat $C(T)$. The appearance of a linear-in-$T$ regime at low temperature is evident. A prototype of the results is shown in Fig.\ref{fig:tre} for various values of the diffusion constant of the hydrodynamic modes.

When the Debye $\sim T^3$ term is added to the specific heat (i.e. away from the marginal stability point), a crossover between the linear regime and the Debye one appears, as shown in Fig.\ref{fig:new}. The importance of the two terms depends on the strength of the diffusion constant of the quasi-localized states and the speed of sound of the propagating phonon modes. More specifically,  for smaller sound speeds the Debye term, which scales like $\sim v^{-3}$ becomes more and more important and the crossover happens for smaller values of temperature. The same effect is produced by increasing the mobility, \textit{i.e.} diffusion constant, of the diffusons. 
Based on the above analytic arguments, we can easily obtain a scaling estimate of the crossover temperature from the linear-in-$T$ regime to the cubic in $T$ regime
\begin{equation}
T^{*} \sim \sqrt{\frac{v^{3}}{D}}
\end{equation}
in terms of the speed of sound and mode diffusion constant.
In summary, the regime of linear in $T$ specific heat extends towards larger temperature the bigger localization of the diffusons (the lower their diffusion constant) and the bigger the speed of sounds in the material.
\begin{figure}[h!]
\centering
\includegraphics[width=0.95\linewidth]{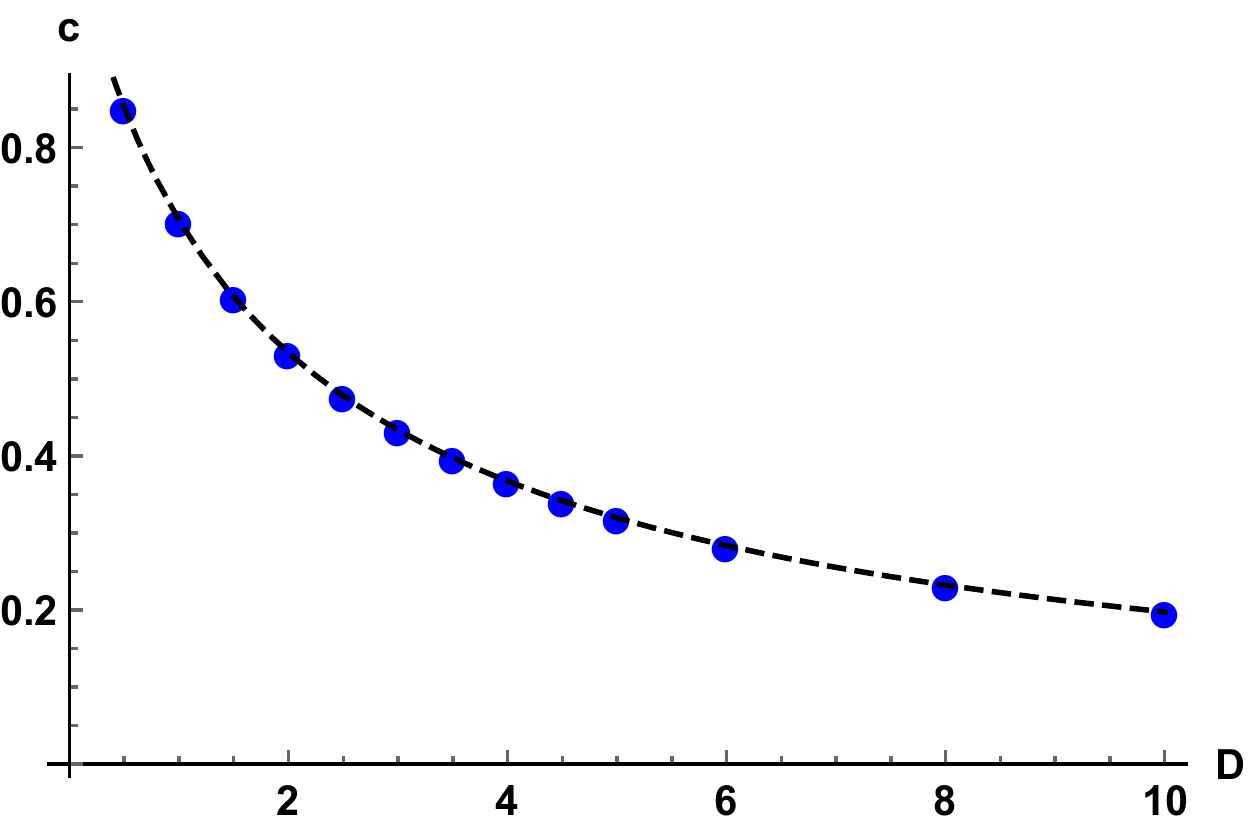}
\caption{The dependence of the linear-in-$T$ coefficient $c$ in the specific heat upon the diffusion constant $D$ of the diffusons. The more the modes are localized, the bigger the linear-in-$T$ contribution. The blue dots are the numerical data while the dashed curve comes from using the analytic expression Eq.\eqref{analT} together with Eq. \eqref{ff}.}
\label{fig:quattro}
\end{figure}\\

Finally, we discuss the dependence of the linear in $T$ term in the specific heat upon varying the diffusion constant $D$ of the diffusons. We obtain an analytic results in Eq.\eqref{analT}, combined with Eq.\eqref{ff}, which is in perfect agreement with the numerical data as shown in Fig.\ref{fig:quattro}.
The higher the diffusion constant of the hydrodynamic modes the smaller their linear in $T$ contribution to the specific heat. In simple words, we can state that the more localized the modes, \textit{i.e.} the smaller their diffusion constant $D$, the more important the linear-in-$T$ contribution to the specific heat. This last observation suggests a strong correlation between the degree of localization of the low energy vibrational modes and the anomalous linear-in-$T$ scaling ubiquitously observed in glasses and amorphous solids.\\


We built a simple and analytic effective-field theory model to explain the low-$k$/low-frequency anomalous properties of glasses and amorphous materials. The framework is based on hydrodynamic arguments and on the introduction of new quasilocalized diffusive modes, \textit{i.e.} the diffusons, which crucially modify the dynamics at low energy, and become dominant in the limit of marginally-stable solids where transverse modes and the shear modulus vanish.
We directly show that these modes induce a constant-in-frequency plateau in the vibrational density of states (VDOS) at low frequency. Importantly, these modes produce a linear in $T$ term in the specific heat which is widely observed in experiments on highly disordered systems like glasses.
Also, we analytically derived the relation \eqref{analT} between the coefficient of the linear term in the specific heat and the diffusion constant of the diffusons, which establishes an unprecedented link between the linear-in-$T$ specific heat and the degree of (quasi)localization of vibrational modes in disordered systems.

Our study sheds light on the correlation between disorder, hydrodynamics and eigenmode diffusion. 
It is tempting to indicate simple mode-diffusive dynamics as the underlying universal reason for all the known ''anomalous" properties of disordered solids (specific heat, thermal conductivity, vibrational spectra). Our effective field theory approach does not rely on any specific microscopic mechanism, but it would certainly be desirable to have a more precise understanding of these diffusive quasilocalized modes in relation to the microscopic scattering processes induced by the disorder.

This description can be used in the future also in an attempt to arrive at a semi-analytical theory of thermal conductivity in disordered materials as well as in crystalline materials where anharmonicity induces scattering and quasi-localization of modes~\cite{Akkermans}, possibly in combination with the Allen-Feldman framework~\cite{allen1999diffusons}. 

Finally, it would be interesting to verify if the diffusive dynamics and the correlated properties discussed in this letter might be at work also in other systems in the low-$k$ hydrodynamic regime~\cite{Sciortino}. A possible playground to consider is the dynamics of the viscous electrons in graphene and Dirac materials in the hydrodynamic window \cite{Lucas:2017idv}. Also, this framework may be relevant to network-forming glasses where marginally-stable (or floppy) mesoscopic regions are present along with stable (rigid)ones~\cite{Ferrante2013}.

Moreover, it would be important to find more signatures for the presence of these diffusive modes, looking for example at other experimental observables, such as transport coefficients and conductivities. A ballistic to hydrodynamic crossover is expected at the point of marginal stability and it might induce more effects than the ones discussed in this manuscript.

\section*{Acknowledgments}
We thank Silvio Franz and Richard Stephens for useful comments and discussions about the topics considered in this work. 
M.B. acknowledges the support of the Spanish Agencia Estatal de Investigacion through the grant “IFT Centro de Excelencia Severo Ochoa SEV-2016-0597.
M.B. thanks the University of Milano for the hospitality during the initial stages of this work.
\medskip
 
\bibliographystyle{unsrt}
\bibliography{hydrodiffusons}

\end{document}